\documentclass[12pt]{article}
\usepackage{color}
\usepackage{amssymb,latexsym,bm,amsmath,amsthm,amsfonts,multirow,graphicx,mathrsfs}
\usepackage{algorithm}
\usepackage{algorithmic}
\usepackage{epsfig}
\usepackage{enumitem}

\newcommand{\argmin}{\mathop{\rm argmin}}

\graphicspath{{fig/}}

\newtheorem{thm}{Theorem}[section]

\newtheorem{rmk}[thm]{Remark}

\def\vec{{\rm vec}}
\newcommand{\rmspan}{{\rm span}}

\def\e{\mathcal{E}}

\def\senv{\mathcal{S}_{\rm env}}

\def\denv{d_{\rm env}}

\def\A{\mathcal{A}}
\def\S{\mathcal{S}}

\def\syx{\mathcal{S}_{Y|X}}

\newcommand{\indep}{\;\, \rule[0em]{.03em}{.67em} \hspace{-.25em}
\rule[0em]{.65em}{.03em} \hspace{-.25em}
\rule[0em]{.03em}{.67em}\;\,}

\title{Sufficient Dimension Reduction via Random-Partitions
for Large-$p$-Small-$n$ Problem}
\author{}
\date{}

\begin{document}

\author{Hung Hung$^{1}$
and Su-Yun Huang$^{2}$\\[2ex]
\small $^1$Institute of Epidemiology and Preventive Medicine,
National
Taiwan University, Taiwan\\
\small $^2$Institute of Statistical Science, Academia Sinica,
Taiwan}

\maketitle

\begin{abstract}
Sufficient dimension reduction (SDR) is continuing an active
research field nowadays for high dimensional data. It aims to
estimate the central subspace (CS) without making distributional
assumption. To overcome the large-$p$-small-$n$ problem we propose a
new approach for SDR. Our method combines the following ideas for
high dimensional data analysis: (1)~Randomly partition the
covariates into subsets and use distance correlation (DC) to
construct a sketch of envelope subspace with low dimension.
(2)~Obtain a sketch of the CS by applying conventional SDR method
within the constructed envelope subspace. (3)~Repeat the above two
steps for a few times and integrate these multiple sketches to form
the final estimate of the CS. We name the proposed SDR procedure
``integrated random-partition SDR (iRP-SDR)''. Comparing with
existing methods, iRP-SDR is less affected by the selection of
tuning parameters. Moreover, the estimation procedure of iRP-SDR
does not involve the determination of the structural dimension until
at the last stage, which makes the method more robust in a
high-dimensional setting. Asymptotic properties of iRP-SDR are also
established. The advantageous performance of the proposed method is
demonstrated via simulation studies and the EEG data analysis.

.

~\

\end{abstract}

\clearpage

\section{Introduction}\label{sec.1}

In many modern applications, the number of covariates are often too
large to provide a parsimonious interpretation or to have insights
into the data set. Sufficient dimension reduction (SDR) is thus
continuing an active research field nowadays. Let $Y\in \mathbb{R}$
be the response of interest, and let $X=(X_1,\ldots,X_p)^\top\in
\mathbb{R}^p$ be the covariates with $E(X)=0$ and ${\rm
cov}(X)=\Sigma$. SDR aims to search for a subspace of $\mathbb{R}^p$
with a basis $B$ such that
\begin{eqnarray}
Y\indep X |B^\top X.  \label{SDR}
\end{eqnarray}
The intersections of all such $\rmspan(B)$ exists under certain
conditions (Cook, 1994), which is called the central subspace (CS)
for the regression of $Y$ on $X$, and is denoted by $\syx$ with the
structural dimension $d={\rm dim}(\syx)$. With $B$ being obtained,
subsequent analysis can be based on the lower dimensional $(Y,B^\top
X)$ without losing information. Since the pioneering work of Li
(1991), there are many methods developed to estimate $\syx$. One
branch of SDR methods can be formulated as the following eigenvalue
problem
\begin{eqnarray}\label{sdr.criterion}
K\beta_j=\lambda_j \beta_j\quad{\rm with}\quad K=\Sigma^{-1}M,\quad
j=1,\ldots,p,
\end{eqnarray}
where $M$ is a method-specific symmetric matrix. The leading $d$
eigenvectors $\beta_j$'s (normalized to $\beta_j^\top\Sigma
\beta_j=1$) then provide an estimate of $\syx$. For example, the
sliced inverse regression (SIR, Li, 1991) uses $M={\rm
cov}\{E(X|Y)\}$ and the sliced average variance estimation (SAVE,
Cook and Weisberg, 1991) uses $M=\Sigma^{1/2}E[\{I-{\rm
cov}(\Sigma^{-1/2}X|Y)\}^2]\Sigma^{1/2}$. We refer the reader to Ma
and Zhu (2013) for a review of SDR methods.

Most of the conventional SDR methods become unstable when $p\approx
n$ or even fail to apply when $p\gg n$, due to the matrix inversion
$\Sigma^{-1}$ in~(\ref{sdr.criterion}). This drawback has limited
the usage of many SDR methods when $p$ is large. The problem of
inverting $\Sigma$ can be avoided if we can find an {\it envelope
subspace} $\senv$ such that
\begin{eqnarray}
\syx\subseteq\senv,   \label{env}
\end{eqnarray}
whose dimension $ \denv={\rm dim}(\senv)$ is relatively smaller than
$n$. Let $\e\in \mathbb{R}^{p\times \denv}$ be an orthonormal basis
of $\senv$, then (\ref{env}) implies the existence of a matrix
$\Gamma\in \mathbb{R}^{\denv\times d}$ such that the basis $B$ of
$\syx$ can be expressed as
\begin{eqnarray}
B=\e \,\Gamma.  \label{e.B}
\end{eqnarray}
As a result, we have from (\ref{SDR}) that $Y\indep \e^\top X\,|\,
\Gamma^\top(\e^\top X)$, which gives $\rmspan(\Gamma)=\S_{Y|\e^\top
X}$. One can then apply any SDR method on $(Y,\e^\top X)$ to
estimate $\Gamma$ (which is doable since $\denv < n$), and transform
back to $\mathbb{R}^p$ via (\ref{e.B}) to estimate $\syx$. For
instance, a commonly used strategy to deal with the
large-$p$-small-$n$ problem is to apply SDR methods after PCA
(PCA-SDR), which is equivalent to construct $\e$ by the leading
eigenvectors of $\Sigma$. A similar idea can also be found in the
partial inverse regression estimate (PIRE) of Li, Cook and Tsai
(2007) and the seeded dimension reduction of Cook, Li and
Chiaromonte (2007), where $\e$ is constructed by Krylov sequence.

Different from the above-mentioned methods of PCA-SDR or PIRE, there
are another branch of SDR methods, which handle the
large-$p$-small-$n$ problem by (i) conducting SDR methods in many
lower dimensional subspaces and (ii) integrating SDR results from
these sub-problems to obtain a final SDR analysis. Let $\Pi \in
\mathbb{R}^{p\times p}$ be a column permutation matrix when
multiplied on the right side of a matrix. A size-$r$ subset of $X$
can be written as $\Omega_r^\top X =: X_{\Omega_r}$, where
$\Omega_r$ is a \emph{sampling matrix} given by
\begin{eqnarray}
\Omega_r^\top = \left[ I_{r\times r},\, {\bm
0}_{r\times(p-r)}\right]\Pi. \label{O}
\end{eqnarray}
For SDR on $(Y,X_{\Omega_r})$, one only needs to invert
$\Omega_r^\top\Sigma \Omega_r$, which is of size $r\times r$. Hilafu
(2015) proposed randomized SIR (rSIR) by repeatedly applying SIR on
$(Y,  X_{\Omega_r})$ with multiple randomly generated $\Omega_r$'s.

The idea of reducing the problem size by analyzing a random
sub-problem has been discussed in the literature of randomized
numerical linear algebra and random sketching (see, e.g., Halko,
Martinsson and Tropp, 2011; Woodruff, 2014), where a sub-problem
provides a sketch of the original problem. The idea of integrating
results from multiple sub-problems to improve the accuracy of data
analysis can also be found in literature. Chernoff, Lo and Zheng
(2009) have demonstrated that influential variables can be
effectively identified by repeatedly inspecting the association
between $Y$ and $ X_{\Omega_r}$ for multiple random sampling
matrices $\Omega_r$'s. Li, Wen and Zhu (2008) proposed to integrate
SDR results from multiple random projections of $Y$ to estimate
$\syx$ when $Y$ is multivariate. Chen {\it et al.} (2016) proposed
to integrate multiple random sketches of singular value
decomposition. On the other hand, properly using the concept of
$\senv$ to confine the inferential target not only can enhance the
estimation efficiency, but also can increase the interpretability of
analysis results. The aim of this work is to propose a new SDR
method, called ``integrated random-partition SDR (iRP-SDR)'', to
deal with the large-$p$-small-$n$ problem, by utilizing both the
ideas of envelope subspace $\senv$ and integration of results from
multiple random subsets $ X_{\Omega_r}$'s.

\section{Method: Integrated Random-Partition SDR}

The proposed iRP-SDR consists of three major steps:
\begin{enumerate}\itemsep=0pt
\item
A sketch of the envelope subspace $\senv$ is constructed by a
combination of random-partition and distance-correlation
screening (see Section~\ref{sec.2.senv}).

\item
Conventional SDR method is applied within the sketch of $\senv$
to estimate the kernel matrix $K$ in (\ref{sdr.criterion}) that
avoids inverting $\Sigma$ (see Introduction). Note that the
estimate of the kernel matrix depends on the random-partition.

\item
Steps~1-2 are repeated a few times and the resulting kernel
matrices are integrated. Multiple runs together with an
integration to form the final estimate of $\syx$ can reduce the
variation due to random-partitions. (see
Section~\ref{sec.2.est}).
\end{enumerate}

\subsection{A sketch of $\bm{\senv}$ via random-partition
and DC screening}\label{sec.2.senv}

Define the active set of $X$ to be
\begin{eqnarray}
\A=\Big\{X_j:B(j,k)\ne 0~{\rm for~some}~k\Big\} \label{A}
\end{eqnarray}
with $B(j,k)$ being the $(j,k)$-th element of $B$ in~(\ref{SDR}).
The active set $\A$ contains the elements of $X$ that appear in the
conditional distribution of $Y$ given $X$. Assume $|\A|<\infty$ and
let
\begin{eqnarray}
\S_\A=\rmspan\{e_j:X_j\in \A\},
\end{eqnarray}
where $e_j\in \mathbb{R}^p$ is the vector with 1 in the $j$-th place
and 0 elsewhere. Certainly, $\syx\subseteq\S_\A$, and any space
containing $\S_\A$ can serve as an envelope subspace $\senv$
fulfilling (\ref{env}). The construction of $\senv$ can then be
achieved by a proper estimation of $\A$. Fan and Lv (2008) proposed
sure independence screening (SIS) to estimate $\A$ by retaining
$X_j$'s with leading absolute values of Pearson correlation
coefficients with $Y$. They showed that SIS possesses the \emph{sure
screening property} under the linear regression model for $Y$ given
$X$. The linear regression assumption, however, can be violated in
some situations. Moreover, SIS in its nature is a marginal screening
method, which ignores the joint effects among $X$. To take the joint
effects among $X$ into account, we adopt a screening method based on
the distance correlation (DC, Szekely, Rizzo and Bakirov, 2007) to
recover $\A$. The squared DC between two random vectors $(v_1,v_2)$
is defined to be
\begin{eqnarray}
\omega_{\rm dc}(v_1,v_2)=\frac{{\rm dcov}^2(v_1,v_2)}{{\rm
dcov}(v_1,v_1) {\rm dcov}(v_2,v_2)},
\end{eqnarray}
where
\begin{eqnarray*}
{\rm dcov}^2(v_1,v_2)&=&
E\left\{\|v_1-\widetilde{v_1}\|\cdot\|v_2-\widetilde{v_2}\|\right\}
+E\|v_1-\widetilde{v_1}\|\cdot E\|v_2-\widetilde{v_2}\|\nonumber\\
&-&2E\left\{E(\|v_1-\widetilde{v_1}\| | v_1)\cdot
E(\|v_1-\widetilde{v_2}\||v_2)\right\}
\end{eqnarray*}
and $(\widetilde{v_1},\widetilde{v_2})$ is a random copy of
$(v_1,v_2)$. The reasons of using DC are threefold. First, DC
measures a general association between two random vectors
$(v_1,v_2)$, in the sense that $v_1\indep v_2$ if and only if
$\omega_{\rm dc}(v_1,v_2)=0$. Second, DC is induced from the
characteristic function, which is totally model-free. Third, DC can
be applied to cases where $v_1$ and $v_2$ are not of the same
dimension, which is able to measure the association between $Y$ and
a subset $X_{\Omega_r}$.

To recover $\A$ via using $\omega_{\rm dc}$, define a \emph{size-$r$
random-partition} of $X$ to be a collection of sampling matrices in
(\ref{O}):
\begin{eqnarray}
\mathcal{P}_r=\left\{\Omega_{r,k}: k=1,\ldots,\frac{p}{r}\right\}
\mbox{~such that~$\{ X_{\Omega_{r,k}}\}_{k=1}^{p/r}$ forms a partition of
$X$.}
\end{eqnarray}
For simplicity, we assume that $p/r$ is an integer in the rest of
discussions. For the case of general $p$, there are $\lfloor
p/r\rfloor$ subsets with size $r$ and one subset with size
$p-\lfloor p/r\rfloor r$, where $\lfloor \cdot\rfloor$ denotes
taking the integer part. For a given $\mathcal{P}_r$, we propose to
estimate $\A$ by
\begin{eqnarray}
\widehat\A_u^{(\mathcal{P}_r)} = \Big\{ X_{\Omega_{r,k}}:
 \Omega_{r,k}\in\mathcal{P}_r,
 ~~ \widehat\omega_{\rm dc}(Y, X_{\Omega_{r,k}})\ge c\Big\},\label{Au}
\end{eqnarray}
where $\widehat\omega_{\rm dc}$ is the sample version of
$\omega_{\rm dc}$ by replacing expectations with empirical moment
estimators, $c$ is a critical value, and $u=u(c)$ denotes the
dimension of $\widehat\A_u^{(\mathcal{P}_r)}$ under $c$. The
selection of critical value will be discussed later. Note that, for
$r=1$, $\widehat\A_u^{(\mathcal{P}_r)}$ is exactly the marginal
screening criterion studied by Li, Zhong and Zhu (2012). They also
mentioned the superiority of $\omega_{\rm dc}$ in measuring the
association between $Y$ and a subset of $X$, which motivates us to
estimate $\A$ by using $\widehat\A_u^{(\mathcal{P}_r)}$ with $r\ge
1$. By considering the association between $Y$ and $
X_{\Omega_{r,k}}$'s, it not only can take the joint effects among
$X$ into account (which is able to integrate weak signals in a
subset to a stronger one), but also can reduce the number of units
under consideration (which has the potential to increase the power
of detecting variables in $\A$). Finally, a basis of $\senv$ is
constructed to be the $p\times u$ matrix
\begin{eqnarray}
\widehat\e_u^{(\mathcal{P}_r)}=\big[\,e_{j}:X_j\in
\widehat\A_u^{(\mathcal{P}_r)} \,\big],\label{e.dc}
\end{eqnarray}
which will be used to develop iRP-SDR in the next subsection. Note
that $\widehat\e_u^{(\mathcal{P}_r)}$, and hence the subsequent
analysis, depends on the choices of $\mathcal{P}_r$ and the
dimension~$u$. These issues of stochastic variation in
random-partition and hyper-parameters selection will be discussed in
Sections~\ref{sec.2.est}-\ref{sec.tuning}, respectively.

We close this section by justifying the use of
$\widehat\e_u^{(\mathcal{P}_r)}$ as a sketch of $\senv$. It mainly
relies on the sure screening property of DC screening, which
requires that the probability of $\widehat\A_u^{(\mathcal{P}_r)}$
containing $\A$ approaches one as $n\to\infty$. Li, Zhong and Zhu
(2012) have established the sure screening property for
$\widehat\A_u^{(\mathcal{P}_r)}$ with $r=1$. Same arguments can be
applied to the case of arbitrary $r\ge1$ under the following
assumptions:
\begin{itemize}\itemsep=0pt
\item[(C1)]
There exists a positive constant $s_0$ such that for all $0<s\le
2s_0$,
\[E\left\{\exp(s\|Y\|^2)\right\}<\infty~~ {\rm and}~~
\sup_p\max_{1\le j\le p}E\left\{\exp(s\|X_j\|^2)\right\}<\infty.\]

\item[(C2)]
The DC value of $Y$ and $X_{\Omega_r}$, where $X_{\Omega_r}$
contains at least one active variable, is significantly large in
the sense that, for some constants $\kappa_1>0$ and $0\le
\kappa_2<1/2$,
\[\inf_{\left\{\Omega_r:~X_{\Omega_r} \bigcap\A\ne\emptyset\right\}}
\omega_{\rm dc}(Y, X_{\Omega_r})\ge 2\kappa_1n^{-\kappa_2}.\]
\end{itemize}
Condition (C1) is assumed in Li, Zhong and Zhu (2012), and condition
(C2) is modified to adapt to the case of subset size $r\ge 1$. We
have the following result.

\begin{thm}[sure screening property]\label{thm.sure_screening}
Assume conditions (C1)-(C2), and assume the critical value in
$\widehat\A_u^{(\mathcal{P}_r)}$ is chosen to be
$c=\kappa_1n^{-\kappa_2}$, where $(\kappa_1,\kappa_2)$ are defined
in (C2). Then, we have for any $\mathcal{P}_r$ that
\begin{eqnarray}\label{sure_screening}
\lim_{n \to\infty}
P\left\{\A\subseteq\widehat\A_u^{(\mathcal{P}_r)}\big|\mathcal{P}_r\right\}=1,
\end{eqnarray}
where $p$ can take an order $\log p =o\left(n^{(1-2\kappa_2)/3}
\right)$.
\end{thm}

Theorem~\ref{thm.sure_screening} implies that $\lim_{n\to\infty}
P\{\syx\subseteq \rmspan(\widehat\e_u^{(\mathcal{P}_r)})
\big|{\mathcal P}_r\}=1$. It ensures the estimation of $\syx$ can be
based on $(Y, \widehat\e_u^{(\mathcal{P}_r)\top}X)$. Though the span
of $\widehat\e_u^{(\mathcal{P}_r)}$ can work as an envelope
subspace, its dimension has to be restrained. The dimension of
$\widehat\e_u^{(\mathcal{P}_r)}$ is given by $u$, which could
diverge with $p$ if not properly controlled. To make conventional
SDR methods applicable using
$(Y,\widehat\e_u^{(\mathcal{P}_r)\top}X)$, a simple way is to
require $u<\infty$. A sufficient condition to ensure the finiteness
of $u$ is to further assume (C3) below.
\begin{itemize}\itemsep=0pt
\item[(C3)]
The DC value of $Y$ and $X_{\Omega_r}$, where $X_{\Omega_r}$ does
not contain any active variable, is significantly small in the sense
that
\begin{eqnarray*}
\sup_{\left\{\Omega_r:~ X_{\Omega_r} \bigcap\A=\emptyset\right\}}
\omega_{\rm dc}(Y, X_{\Omega_r})=o(n^{-\kappa_2}),
\end{eqnarray*}
where $\kappa_2$ is defined in (C2).
\end{itemize}
The following result is essential for
Theorem~\ref{thm.consistency.pr} below.
\begin{thm}\label{thm.u}
Assume the conditions in Theorem~\ref{thm.sure_screening} and
condition (C3). Then, we have for any $\mathcal{P}_r$ that
\begin{eqnarray*}
\lim_{n \to\infty} P\left\{\widehat\A_u^{(\mathcal{P}_r)}=
\A^{(\mathcal{P}_r)}\big|\mathcal{P}_r\right\}=1,
\end{eqnarray*}
where $\A^{(\mathcal{P}_r)}=\{ X_{\Omega_{r,k}}:\Omega_{r,k}\in
\mathcal{P}_r, X_{\Omega_{r,k}}\bigcap\A\ne\emptyset\}$ with
$u_0=u_0(\mathcal{P}_r)=|\A^{(\mathcal{P}_r)}|$ satisfying $d\le
u_0<\infty$, and $p$ can take an order $\log p
=o\left(n^{(1-2\kappa_2)/3} \right)$.
\end{thm}

\subsection{Estimation of $\bm{\syx}$} \label{sec.2.est}

Below we introduce our iRP-SDR via using
$\widehat\e_u^{(\mathcal{P}_r)}$ in~(\ref{e.dc}). In the rest of
discussions, we use SIR as the core SDR method to explain the
details of our proposal. Extensions to other SDR methods based on
criterion~(\ref{sdr.criterion}) are straightforward. Given
$\mathcal{P}_r$, Theorem~\ref{thm.sure_screening} ensures
\begin{eqnarray}\label{est.relation}
\lim_{n\to\infty}P\left\{B=P_{
\widehat\e_u^{(\mathcal{P}_r)}}B\big|\mathcal{P}_r\right\}= 1,
\end{eqnarray}
where $B$ is a basis of $\syx$, and $P_M$ is the orthogonal
projection matrix onto $\rmspan(M)$ for a matrix $M$. It enables the
estimation of $\syx$ to be based on $(Y,
\widehat\e_u^{(\mathcal{P}_r)\top}X)$, which avoids inverting
$\Sigma$. In particular, let $\widehat\gamma_j^{(\mathcal{P}_r)}$
and $\widehat\lambda_j^{(\mathcal{P}_r)}$ be the $j$-th eigenvector
and eigenvalue obtained from applying SIR on $(Y,
\widehat\e_u^{(\mathcal{P}_r)\top}X)$, $j=1,\ldots,u$. Let also
\begin{eqnarray}
\widehat\beta_j^{(\mathcal{P}_{r})}=\widehat\e_u^{(\mathcal{P}_{r})}
\, \widehat\gamma_j^{(\mathcal{P}_{r})},\quad j=1,\ldots,u,
\label{Bj.s}
\end{eqnarray}
which transforms $\widehat\gamma_j^{(\mathcal{P}_{r})}$ back to
$\mathbb{R}^p$ via~(\ref{e.B}). Since the leading eigenvectors
$\widehat\gamma_j^{(\mathcal{P}_{r})}$'s provide an estimate of
$\mathcal{S}_{Y|\widehat\e_u^{(\mathcal{P}_r)\top}X}$, it implies
from (\ref{est.relation}) that
$\widehat\beta_j^{(\mathcal{P}_{r})}$'s provide an estimate of
$\syx$. The projection matrix (with respect to the $\Sigma$-inner
product) associated with $\widehat\beta_j^{(\mathcal{P}_{r})}$ is
given by
$\widehat\beta_j^{(\mathcal{P}_{r})}\big\{\widehat\beta_j^{(\mathcal{P}_{r})\top}\Sigma
\widehat\beta_j^{(\mathcal{P}_{r})}\big\}^{-1}
\widehat\beta_j^{(\mathcal{P}_{r})\top} \Sigma$, which is estimated
by $\widehat\beta_j^{(\mathcal{P}_{r})}
\widehat\beta_j^{(\mathcal{P}_{r})\top} \widehat\Sigma$. The
projection matrix enables us to summarize the
$\mathcal{P}_{r}$-analysis via the kernel matrix
\begin{eqnarray}
\widehat K_{u}^{(\mathcal{P}_{r})}=\sum_{j=1}^u
\widehat\lambda_j^{(\mathcal{P}_{r})} \,
\widehat\beta_j^{(\mathcal{P}_{r})}\beta_j^{(\mathcal{P}_{r}){\top}}{\widehat\Sigma},
\label{Kr.j}
\end{eqnarray}
where the subscript $u$ indicates that the construction of $\widehat
K_{u}^{(\mathcal{P}_r)}$ is based on the $u$-dimensional
$\widehat\e_u^{(\mathcal{P}_r)}$. One can treat $\widehat
K_{u}^{(\mathcal{P}_r)}$ as an estimate of the SIR kernel matrix
pre-multiplied by $\Sigma^{-1}$, which has
$\widehat\beta_j^{(\mathcal{P}_r)}$ as its eigenvector with
eigenvalue $\widehat\lambda_j^{(\mathcal{P}_r)}$. See
Remark~\ref{rmk.spectral.theory} for details.

Although the kernel matrix $\widehat K_{u}^{(\mathcal{P}_r)}$
provides a basis to estimate $\syx$ without inverting $\Sigma$, it
only produces a sketch estimate with less precision. Moreover, the
analysis result will depend on the choice of the random-partition
$\mathcal{P}_r$. There generally exists no prior knowledge about how
$X$ should be partitioned in the SDR problem. A natural strategy is
to consider the expected value of $\widehat K_{u}^{(\mathcal{P}_r)}$
with respect to the uniform distribution for $\mathcal{P}_r$. An
integrated kernel matrix is proposed to be
\begin{eqnarray}
\widehat K_{u,r} =E_{\mathcal{P}_r}[\widehat
K_{u}^{(\mathcal{P}_r)}] = \frac{1}{N_r}\sum_{l=1}^{N_r}\widehat
K_{u}^{(\mathcal{P}_{r,l})}, \label{Kur}
\end{eqnarray}
where $\{\mathcal{P}_{r,l}:l=1,\ldots,N_r\}$ denotes the collection
of all possible size-$r$ random-partitions of $X$. Plugging in
$K=\widehat K_{u,r}$ to~(\ref{sdr.criterion}), a basis of $\syx$ can
be estimated by the leading $d$ eigenvectors of $\widehat K_{u,r}$.
The consistency of $\widehat K_{u,r}$ is stated below.

\begin{thm}[consistency]\label{thm.consistency.pr}
Assume conditions (C1)-(C3), and assume that the critical value in
$\widehat\A_u^{(\mathcal{P}_r)}$ is chosen to be
$c=\kappa_1n^{-\kappa_2}$, where $(\kappa_1,\kappa_2)$ are defined
in (C2). Assume also that SIR based on $(Y,X_{\Omega})$ is a
consistent estimator of $\mathcal{S}_{Y|X_{\Omega}}$ for any
sampling matrix $\Omega$ satisfying $\syx\subseteq\rmspan(\Omega)$.
Then, as $n\to\infty$, we have $\|P_{\widehat K_{u,r}}-P_B\|_F\to 0$
in probability, where $\|\,\cdot\,\|_F$ stands for matrix Frobenius
norm.
\end{thm}

\begin{rmk}\label{rmk.spectral.theory}
The kernel matrix $K$ in criterion (\ref{sdr.criterion}) is
symmetric in the metric of $\Sigma$, i.e., $\Sigma K$ is symmetric
(Tyler, 1981). The spectral theory then implies that
$K=\sum_j\lambda_j \beta_j\beta_j^\top\Sigma$ with
$(\beta_j,\lambda_j)$ being its eigenvector and eigenvalue. This
representation motivates the construction of $\widehat
K_{u}^{(\mathcal{P}_r)}$ in (\ref{Kr.j}). In this viewpoint,
$\widehat K_{u}^{(\mathcal{P}_r)}$ can be treated as a sketch
estimate of the SIR kernel matrix $\Sigma^{-1}{\rm cov}(E(X|Y))$ but
without the need of inverting $\Sigma$.
\end{rmk}

\subsection{Tuning parameters and structural dimension}\label{sec.tuning}

There are two tuning parameters involved in iRP-SDR, including the
critical value $c$ for constructing $\widehat\e_u^{(\cdot)}$ and the
subset size $r$ of the random-partition $\mathcal{P}_r$. Note that
choosing $c$ is equivalent to choosing the envelope dimension $u$.
We provide two simple settings, $\widehat K_u$ in~(\ref{Ku}) and
$\widehat K$ in~(\ref{K}) below, for the tuning parameters $(u,r)$.

We first deal with the selection of $r$ with a given $u$. The value
of $r$ determines the subset size of the random-partition used to
construct $\widehat\e_u^{(\cdot)}$. A larger $r$ makes $\omega_{\rm
dc}(Y, X_{\Omega_{r,k}})$ more capable to reflect the joint effects
among $X$, but at the cost of being less efficient in estimating
$\omega_{\rm dc}(Y, X_{\Omega_{r,k}})$ with limited sample size $n$.
There generally exists no prior knowledge of an ideal partition
size, and a natural strategy is to consider all possible choices of
$r$. Let $\mathcal{R}_u$ be the candidate set of choices of~$r$,
which consists of the unique elements of
$\{\lfloor\frac{u}{s}\rfloor:s=1,\ldots,u\}$. E.g., for $u=6$, we
have $\mathcal{R}_u =\{1,2,3,6\}$. For any $r\in \mathcal{R}_u$, we
include in $\widehat\A_u^{(\mathcal{P}_r)}$ those subsets $
X_{\Omega_{r,k}}$'s with the largest $u/r$ values of
$\widehat\omega_{\rm dc}(Y, X_{\Omega_{r,k}})$'s. When $u/r$ is not
an integer, we select $\lfloor u/r \rfloor$ subsets. This procedure
gives $|\widehat\A_u^{(\mathcal{P}_r)}|=r\lfloor u/r \rfloor \le u$.
The integrated kernel matrix over $r\in \mathcal{R}_u$ is given by
\begin{eqnarray}
\widehat K_{u}=\sum_{r\in \mathcal{R}_u} \widehat K_{u,r}.
\label{Ku}
\end{eqnarray}
An estimate of $\syx$ is proposed to be $\widehat B_u$, the leading
$d$ eigenvectors of $\widehat K_u$. This kernel matrix $\widehat
K_u$ simplifies the tuning parameter to just one number $u$, the
dimension of the envelope $\widehat\e_u^{(\cdot)}$. Note that most
of the SDR methods for large-$p$-small-$n$ problem eventually face
the issue of choosing a tuning parameter of certain dimensionality.
For example, PCA-SDR needs to determine the number of the leading
eigenvectors of $\widehat\Sigma$, PIRE needs to determine the
dimension of the Krylov sequence, and both rSIR and seq-SDR need to
determine the subset size to reduce the dimension of $X$
sequentially. While Cook, Li and Chiaromonte (2007) proposed a
testing method to determine the dimension of the Krylov sequence,
there is no theoretical support developed concerning this issue in
rSIR and seq-SDR. Considering the fact that $u$ is the reduced model
size such that SIR can be properly implemented using
$(Y,\widehat\e_u^{(\mathcal{P}_r)\top}X)$ with sample size $n$, we
can use $u=\lfloor na\rfloor$ for some $a\in(0,1)$. Of course the
selection of $u$ will affect the performance of iRP-SDR. The optimal
selection of $u$ depends on the underlying data generating
distribution, and is beyond the scope of this work. Alternatively,
we can make the inference procedure less affected by the selection
of $u$, by using an ensemble approach with the integrated kernel
matrix
\begin{eqnarray}
\widehat K = \sum_{u\in \mathcal{U}} \frac{1}{m_u} \widehat K_u,
\label{K}
\end{eqnarray}
where $\mathcal{U}$ is a pre-determined set of possible values of
$u$, and $m_u$ is the sum of eigenvalues of $\widehat K_u$. Dividing
$m_u$ makes the kernel matrices $\frac{1}{m_u} \widehat K_u$'s from
different envelope sizes comparable. Finally, an ensemble estimate
of $\syx$ is proposed to be $\widehat B$, the leading $d$
eigenvectors of $\widehat K$. Note that both $\widehat K_u$ and
$\widehat K$ are finite sums of $\widehat K_{u,r}$'s. The
consistency of $\widehat B_u$ or $\widehat B$ in estimating $\syx$
is thus a direct consequence of Theorem~\ref{thm.consistency.pr}.

The structural dimension $d$ can be determined by existing methods
based on the kernel matrix $\widehat K_u$ or $\widehat K$. We
suggest using the Bayesian information criterion (BIC) of Zhu {\it
et al.} (2010) to select $d$ by
\begin{eqnarray}
\widehat
d=\argmin_{k=1,\ldots,p}\left\{\frac{n\sum_{j=1}^k\{\ln(\ell_j+1)-\ell_j\}}
{2\sum_{j=1}^p\{\ln(\ell_j+1)-\ell_j\}}-2C_n\frac{k(k-1)}{2p}\right\},
\end{eqnarray}
where $\ell_j$'s represent the eigenvalues of $\widehat K_u$ or
$\widehat K$, and $C_n$ is the user-defined penalty. The consistency
of $\widehat d$ follows from the same argument of Zhu {\it et al.}
(2010) and the consistency of $\widehat K_{u}$ or $\widehat K$,
provided that $C_n/n\to 0$ and $C_n\to\infty$ as $n\to\infty$.


\section{Characteristics of iRP-SDR}\label{sec.compare}

The proposed iRP-SDR possesses some characteristics that make it
more adaptive and stable in estimating $\syx$ under the
high-dimensional setting.
\begin{enumerate}
\item[(A1)]
iRP-SDR is adaptive to various situations. The success of
iRP-SDR in recovering $\syx$ mainly relies on the sure screening
property (\ref{sure_screening}) of a ranking and screening
method, which is satisfied under the   conditions (C1)-(C2).

\item[(A2)]
iRP-SDR does not involve the estimation of the structural dimension
$d$ until at the final stage. Most of SDR methods for
large-$p$-small-$n$ problem require determining the structural
dimension during the estimation process, which may suffer the
problem of instability. iRP-SDR performs as the conventional SDR
methods that determines $d$ from the integrated kernel matrix
$\widehat K_u$ or $\widehat K$.

\item[(A3)]
iRP-SDR is easy to implement. Besides the tuning parameters for
the core SDR method (e.g., the slicing number of SIR), iRP-SDR
only depends on the envelope size $u$. iRP-SDR is also able to
combine with any SDR method with the estimation criterion
(\ref{sdr.criterion}). Moreover, iRP-SDR has the potential to
adapt to extremely large data set, since the calculations of
different $\widehat K_{u}^{(\mathcal{P}_{r,l})}$'s can be in
parallel.
\end{enumerate}

\section{Numerical Studies}\label{sec.numerical}

\subsection{Simulation settings}


Let $\varepsilon\sim N(0,1)$ be the error term. We consider the
following models from the literature.
\begin{enumerate}
\item[(M1)]
(Li, Cook and Tsai, 2007). Set $(n,p)=(100,300)$. Each element of
$X$ is from $U(0,1)$, and $Y = \log(|B^\top
X-4|)+\sigma_0\varepsilon$ with
$B=(-0.5,1,0.5,1,-1,-0.8,0.8,1,0.5,0.75,
\textbf{0}_{p-10}^\top)^\top$.

\item[(M2)]
(Yin and Hilafu, 2015). Set $(n,p)=(200,1000)$. Let $X\sim
N(0,\Sigma)$ with the $(j_1,j_2)$-th element of $\Sigma$ being
$0.5^{|j_1-j_2|}$ and $Y=1+\exp(B^\top X)+\varepsilon$ with
$B=(\textbf{0}_{500}^\top, 1,1,1,1, \textbf{0}_{p-504}^\top)^\top$.

\item[(M3)]
(Hilafu and Yin, 2016). Set $(n,p)=(100,500)$. Let $X\sim
N(0,0.5I_p+0.5\textbf{1}_p\textbf{1}_p^\top)$ and $Y =
0.5\exp(0.75\cdot B^\top X)\varepsilon$ with
$B=(-0.5,1,0.5,1,-1,-0.8,0.8,1,0.5,0.75,
\textbf{0}_{p-10}^\top)^\top$.

\item[(M4)]
(Hilafu and Yin, 2016). Set $(n,p)=(100,500)$. Let $Y\sim U(0,1)$
and $X=\beta_1Y+\beta_2Y^2+0.5E$ with $E\sim
N(0,0.5I_p+0.5\textbf{1}_p\textbf{1}_p^\top)$,
$\beta_1=(0.5,0.75,\textbf{0}_{p-2}^\top)^\top$, and
$\beta_2=(0,0,0.75,0.5,\textbf{0}_{p-4}^\top)^\top$. It gives
$B=(0.5I_p+0.5\textbf{1}_p\textbf{1}_p^\top)^{-1}[\beta_1,\beta_2]$.

\end{enumerate}

\begin{figure}[!ht]\hspace{-0.8cm}
\includegraphics[width=3.6in,height=3.2in]{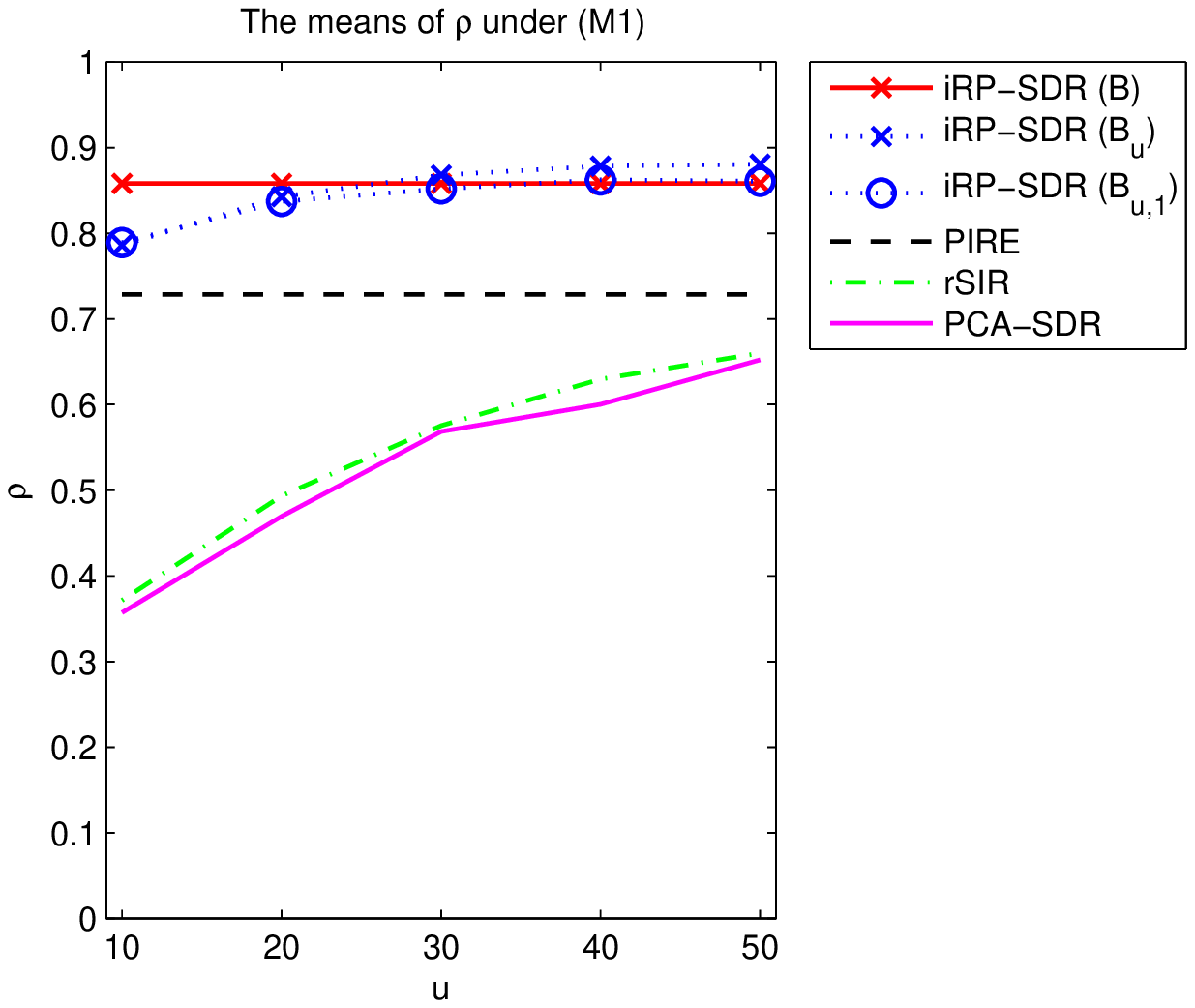}
\includegraphics[width=3.6in,height=3.2in]{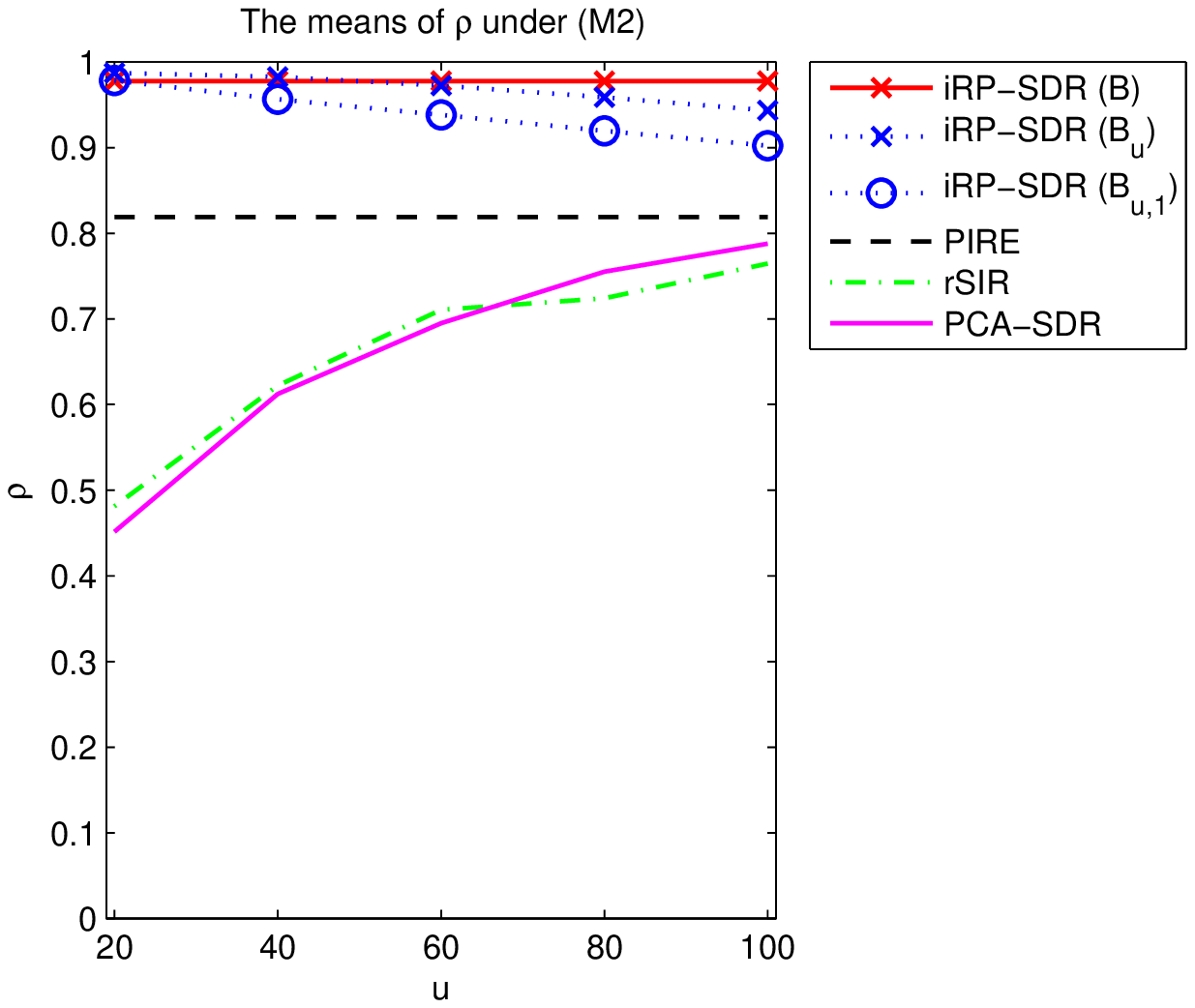}

\hspace{-0.8cm}
\includegraphics[width=3.6in,height=3.2in]{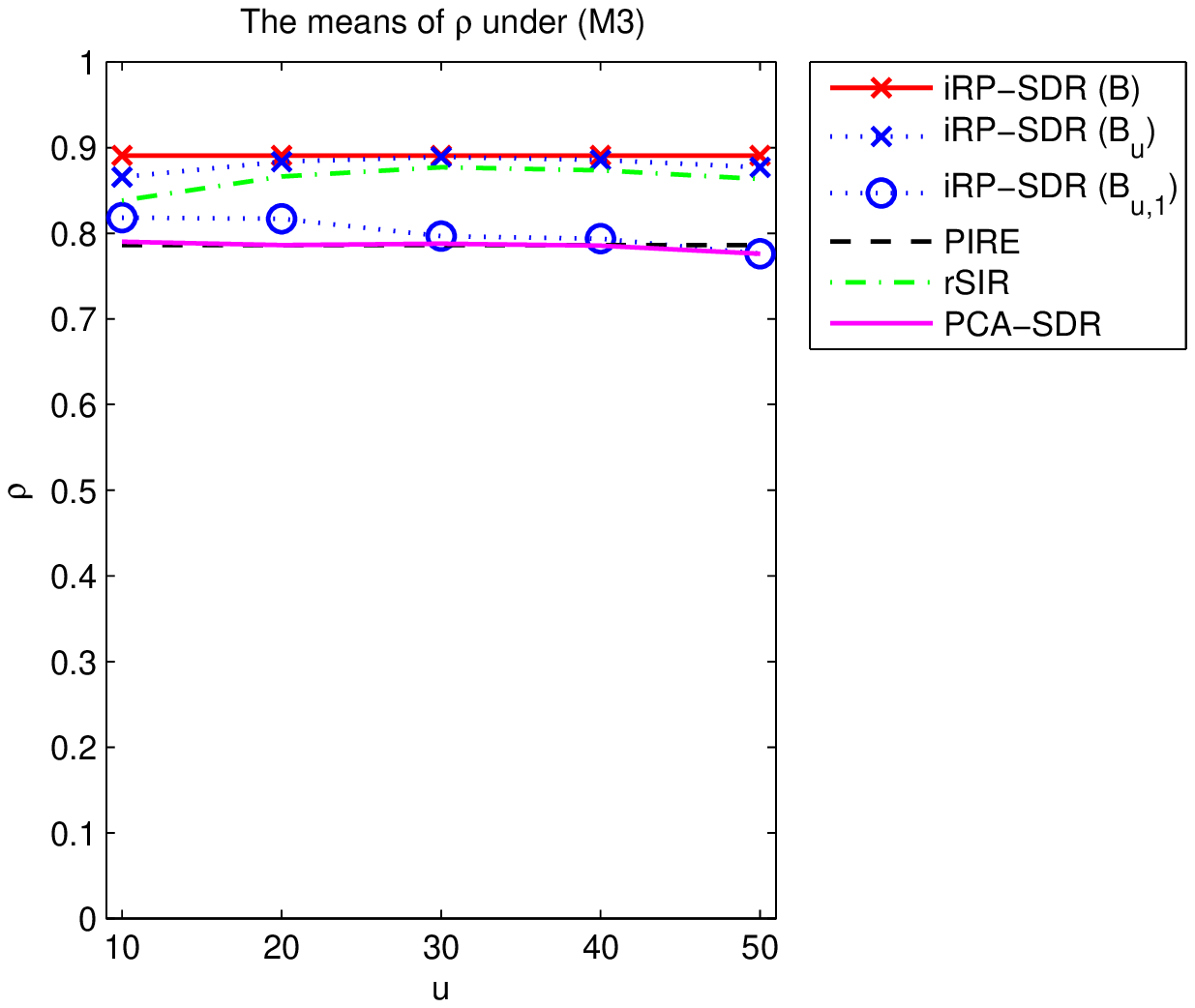}
\includegraphics[width=3.6in,height=3.2in]{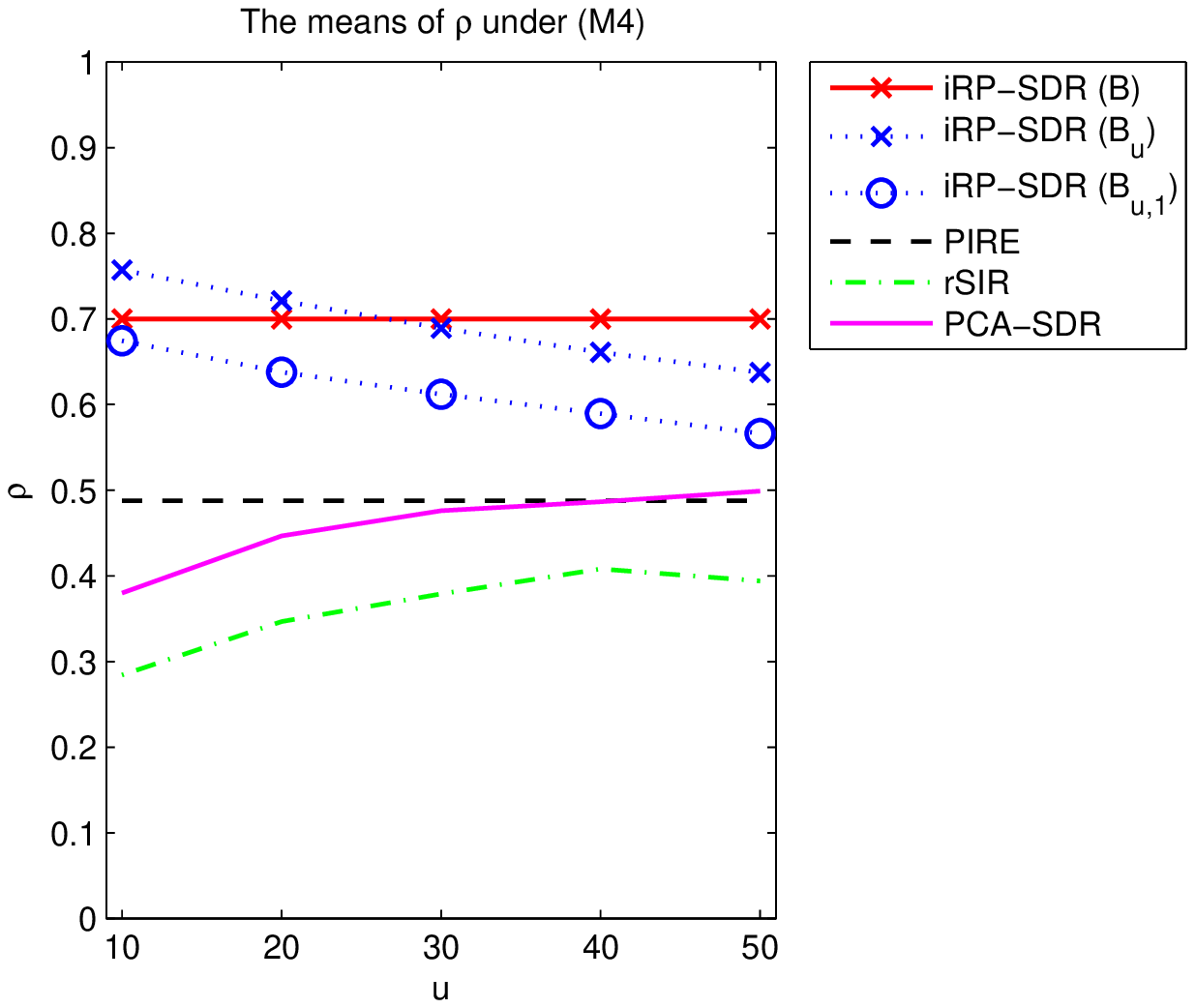}
\caption{The means of $\rho$ at different envelope sizes $u$ under
models (M1)-(M4).}\label{fig.sim1}
\end{figure}

Let $\mathcal{U}=\{0.1n, 0.2n, \ldots, 0.5n\}$ be the candidate set
of the envelope dimension $u$. We implement $\widehat B_u$ of
iRP-SDR with $u\in\mathcal{U}$, PIRE with the Krylov sequence
dimension $u\in \mathcal{U}$, rSIR with the subset size $u\in
\mathcal{U}$, and PCA-SDR with $u\in \mathcal{U}$ leading
eigenvectors of $\widehat\Sigma$, so that all methods use the same
envelope dimension $u$. Following Hilafu and Yin (2016), the slicing
number of SIR used in all methods is set to 5. The mean absolute
value of the trace correlation $\rho=\rho(\widetilde B^\top X,
B^\top X)$ is reported to summarize the performance of an estimator
$\widetilde B$, where $\rho(v_1,v_2)=\sqrt{{\rm trace}(V)/{\rm
dim}(V)}$ with
$V=\Sigma_{v_2}^{-1/2}\Sigma_{v_1v_2}^\top\Sigma_{v_1}^{-1}\Sigma_{v_1v_2}\Sigma_{v_2}^{-1/2}$,
$\Sigma_{v_1}={\rm cov}(v_1)$, $\Sigma_{v_2}={\rm cov}(v_2)$, and
$\Sigma_{v_1v_2}={\rm cov}(v_1,v_2)$, and $\rho=1$ indicates that
$\rmspan(\widetilde B)=\rmspan(B)$. Simulation results of $\rho$
with 100 replicates are placed in Figure~\ref{fig.sim1}. We remind
the reader that all methods considered in our simulation studies use
SIR as the core SDR method. The simulation results then directly
reflect the capability of each method in dealing with the
large-$p$-small-$n$ problem, while controlling the capability of SIR
in estimating $\syx$.

\subsection{Simulation results: comparison with the case of $r=1$}

A critical step of iRP-SDR is the construction of $\senv$ via
random-partitions of $X$ (with subset size $r$) having leading DC
values with $Y$. Recall that using $r>1$ ensures iRP-SDR to take the
joint effects among $X$ into account, while $r=1$ corresponds to
using marginal DC values $\widehat\omega_{\rm dc}(Y,X_j)$'s to
construct $\senv$, which totally ignores the joint effects among
$X$. The first simulation study aims to evaluate the gain from using
$r>1$ to the estimation of $\syx$. To see this, we also report in
Figure~\ref{fig.sim1} the simulation results from the kernel matrix
$\widehat K_{u,r}$ in (\ref{Kur}) with $r=1$ (denoted by $\widehat
B_{u,1}$). Comparing $\widehat B_u$ (from the integrated kernel
matrix $\widehat K_u$) with $\widehat B_{u,1}$, it can be seen that
$\widehat B_u$ outperforms $\widehat B_{u,1}$ uniformly under all
models, especially for the cases of (M2)-(M4). Note that in (M1),
the elements of $X$ are independently generated, under which we gain
less from considering the joint effects among $X$, and $\widehat
B_u$ and $\widehat B_{u,1}$ are detected to have similar
performances. As to (M2)-(M4), $X$ are correlated and the gain from
grouping $X$ becomes obvious. Our simulation study shows the merits
of using $r>1$, and that $\widehat B_u$ outperforms $\widehat
B_{u,1}$ even when the covariates are mutually independent.

\subsection{Simulation results: comparison with other methods}

We first compare $\widehat B_u$ of iRP-SDR with the $\senv$-based
SDR methods: PIRE and PCA-SDR. It can be seen that $\widehat B_u$
outperforms PIRE and PCA-SDR under all models. The performance of
PCA-SDR can be heavily affected by the choice of $u$, especially for
the cases of (M1)-(M2). Recall that PCA-SDR assumes that $\syx$ is
spanned by the leading eigenvectors of $\Sigma$. This condition can
be satisfied for a large $u$ only. A large $u$, however, can also
include in $\senv$ more irrelevant directions outside $\syx$, which
further decreases the estimation efficiency. PIRE also requires
$\syx$ to be spanned by the leading directions of Krylov sequence.
Unlike PCA-SDR, the construction of Krylov sequence in PIRE uses the
information of $Y$. However, our simulation results indicate a
limitation of the Krylov sequence in capturing $\syx$, where PIRE
cannot have better performance than $\widehat B_u$ for all $u$.
Recall the validity of iRP-SDR merely relies on the sure screening
property, which is not related to any specific structure of $\syx$.
iRP-SDR is thus expected to be more adaptive to various situations.
Another reason for the unsatisfactory performance of PIRE and
PCA-SDR is that their construction of $\senv$ involves a
$p$-dimensional eigen-decomposition (i.e., eigenvectors of
$\widehat\Sigma$ in PCA-SDR, and $\nu$ in PIRE) with $n\ll p$. On
the other hand, iRP-SDR constructs $\senv$ via random-partitions of
$X$, each with subset size $r$ only. Considering the limited sample
size, it is also reasonable to expect an efficiency gain of iRP-SDR
over PIRE and PCA-SDR. We next compare $\widehat B_u$ with the
subset-based SDR methods: rSIR. Although rSIR has comparable
performances with $\widehat B_u$ under (M3), it fails to identify
$\syx$ under (M1), (M2), and (M4). It indicates that simply using
random subset of $X$ cannot provide a consistent estimate of $\syx$,
and the naive integration method is not suitable to integrate
multiple results, either.

The simulation results of the ensemble approach $\widehat B$ over
$u\in\mathcal{U}$ are also reported in Figure~\ref{fig.sim1}. It can
be seen that $\widehat B$ always produces comparable results with
$\widehat B_u$, and also dominates other competitors. It implies
that $\widehat B$ is less affected by the selection of the envelope
size and can achieve satisfactory results. Thus, the ensemble
$\widehat B$ is suggested in practice.


\section{The EEG Data}

The EEG data set (downloaded from the \emph{UCI machine learning
repository}) consists of $n=122$ samples, each with a 256$\times$64
matrix $X_0$ and a binary alcoholic status $Y$. The $(j,k)$-th
element of $X_0$ represents the voltage value of the $k$-th probe
measured at the $j$-th time point. It is of interest to construct a
prediction model based on the voltage value for the alcoholic
status. In our analysis, we preprocess the data matrix $X_0$ to form
$\bar X_0$, where $\bar X_0(j,k)={\rm
median}\{X_0(\ell,k):32(j-1)+1\le \ell  \le 32j\}$, $k=1,\ldots,64$.
That is, $\bar X_0$ is obtained from summarizing $X_0$ over $32$
time points, while keeping the data structure of 64 channels. It
gives the dimension of $\bar X_0$ to be $8 \times 64$, where the
$(j,k)$-th element of $\bar X_0$ represents the median voltage value
of the $k$-th probe over the time period $[32(j-1)+1, 32j]$. We then
use the data $(Y,X)$ to enter our analysis, where $X=\vec(\bar X_0)$
has dimension $p=8\times 64 = 512$.

\begin{figure}[!ht]
\begin{center}
\includegraphics[width=5in]{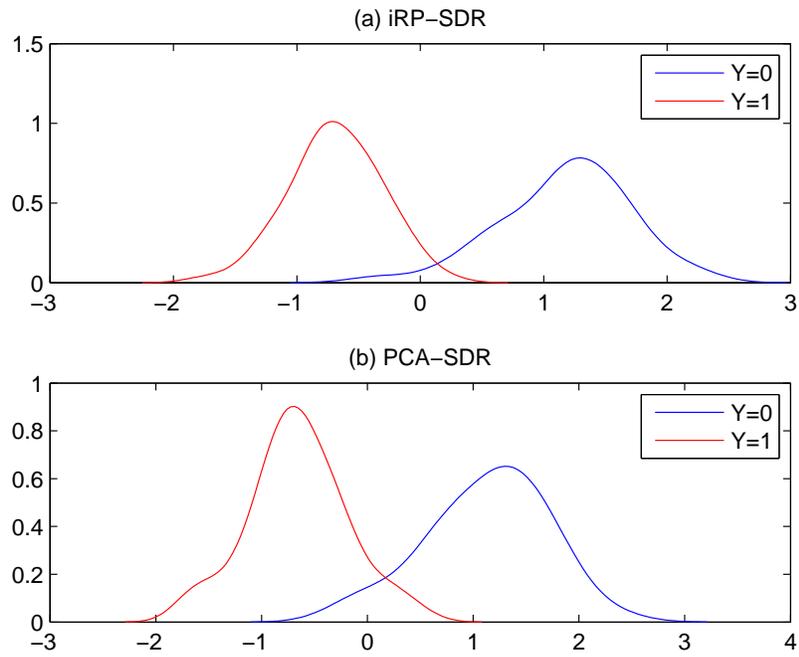}
\end{center}
\caption{The density estimates for two groups of subjects in the EEG
data analysis: (a)~iRP-SDR ($\widehat B_{50}$); and (b) PCA-SDR
($u=50$).}\label{fig.eeg.density}
\end{figure}

\begin{figure}[!ht]
\begin{center}
\includegraphics[width=5in]{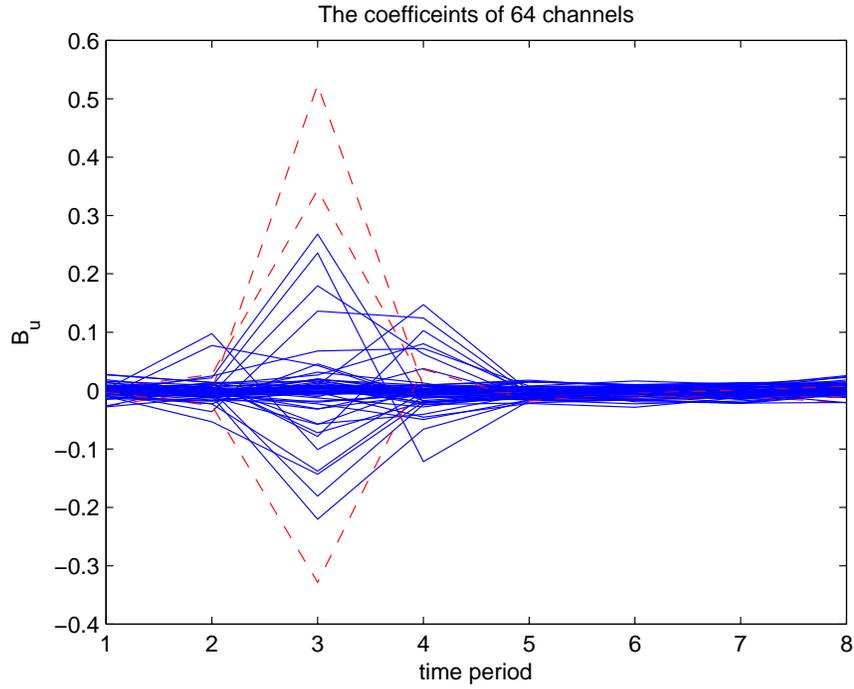}
\end{center}
\caption{The coefficients $\widehat B_{50}$ in the EEG data
analysis, where each of the 64 curves represents the coefficients of
a channel at 8 time periods. The dotted curves represent the
channels O2 ($k=30$), P2 ($k=60$), and P4 ($k=24$) that have
absolute values of $\widehat B_{50}$ larger than 0.3 at the 3rd time
period, i.e., $|\widehat B_{50}(3,k)|>0.3$.}\label{fig.eeg.Bu}
\end{figure}

The analysis result of iRP-SDR ($\widehat B_u$) with $u=50$ and
$d=1$ (since SIR can identify a single direction for binary $Y$) is
placed in Figure~\ref{fig.eeg.density} (a), which reports the
density estimates of $\widehat B_{50}^\top X|Y=y$, $y=0,1$. One can
see that the two density estimates from iRP-SDR are well separated,
which indicates a clear separation on the means of the EEG signals
for two types of subjects. The analysis result of PCA-SDR (with
$u=50$) is reported in Figure~\ref{fig.eeg.density} (b) for
comparison. Although the density estimates from PCA-SDR also show a
clear separation of locations, the overlapping area is detected to
be larger than that from iRP-SDR. It demonstrates the capability of
iRP-SDR to extract more information from high-dimensional data. The
coefficients $\widehat B_{50}$ (from using component-wisely
standardized $X$) from iRP-SDR are reported in
Figure~\ref{fig.eeg.Bu}, where each of the 64 curves represents the
coefficients of a channel at 8 time periods. It is observed that the
curves tend to have a large absolute coefficients at the 3rd time
period, but have nearly zero effects after the 4th time period. It
indicates an early reaction of the brain to the stimulus for
alcoholic patients. The dotted curves represent the channels having
absolute values of $\widehat B_{50}$ larger than 0.3 at the 3rd time
period (i.e., $|\widehat B_{50}(3,k)|>0.3$), including O2 ($k=30$),
P2 ($k=60$), and P4 ($k=24$) that locate nearly on the right brain.
It suggests that the areas of Parietal and Occipital on the right
brain control the reaction to alcoholic stimulus.

To further demonstrate the performance of iRP-SDR, we construct a
prediction model based on $(Y,\widehat B_u^\top X)$ by linear
discriminant analysis (LDA), and the leave-one-out classification
accuracy (CA) from the whole procedure (i.e., SDR followed by LDA
prediction) with different $u$ values are reported in
Table~\ref{tab.eeg}. One can see that iRP-SDR produces higher CA
values than PCA-SDR for every envelope size $u$. Moreover, the
performances of iRP-SDR are quite stable for different choices of
$u$. Our EEG data analysis again demonstrates the superiority of
iRP-SDR in estimating $\syx$ when $n\ll p$.


\section{Discussion}

In this paper, we propose a novel iRP-SDR method for
large-$p$-small-$n$ SDR problem. The superiority of iRP-SDR comes
from the combination of $\senv$ and random-partition as well as
integration of results from multiple random-partitions. The
construction of $\senv$ ensures the consistency of iRP-SDR in
identifying $\syx$, while the random-partition makes iRP-SDR to take
the joint effects among $X$ into account. iRP-SDR is also easy to
implement with a single tuning parameter of the envelope size $u$,
and the computation of $\widehat K_{u}^{(\mathcal{P}_{r,l})}$ can be
put in parallel. The superiority of iRP-SDR is demonstrated via
numerical studies and the EEG data set.

In iRP-SDR, we use DC as the ranking method to construct $\senv$.
There exist other ranking methods that are able to measure the
association between $Y$ and a subset of $X$. We note that any
ranking method satisfying the sure screening property
(\ref{sure_screening}) can be used in iRP-SDR. Another feature that
could affect the performance of iRP-SDR is the integration method
for multiple results. In this paper, we use sample mean to form the
integrated kernel matrix (\ref{Kur}) for simplicity. It is of
interest to study the effects of different ranking and integration
methods on the performance of iRP-SDR.

\section*{References}
\begin{description}
\item
Chen, T. L., Chang, D., Huang, S. Y., Chen, H., Chang, C. and
Wang, W. (2016). Integrating multiple random sketches for
singular value decomposition. arXiv:1608.08285.

\item
Chernoff, H., Lo, S. H., and Zheng, T. (2009). Discovering
influential variables: a method of partitions. \emph{The Annals
of Applied Statistics}, 1335-1369.

\item
Cook, R. D. (1994). On the interpretation of regression plots.
\emph{Journal of the American Statistical Association}, 89(425),
177-189.

\item
Cook, R. D., Li, B., and Chiaromonte, F. (2007). Dimension
reduction in regression without matrix inversion.
\emph{Biometrika}, 94(3), 569-584.

\item
Cook, R. D. and Weisberg, S. (1991). Discussion of ``Sliced
inverse regression for dimension reduction''. {\it Journal of
the American Statistical Association}, 86, 328-332.

\item
Fan, J. and Lv, J. (2008). Sure independence screening for
ultrahigh dimensional feature space. \emph{Journal of the Royal
Statistical Society: Series B}, 70, 849-911.

\item
Halko, N., Martinsson, P. G. and Tropp, J. A. (2011). Finding
structure with randomness: probabilistic algorithms for
constructing approximate matrix decompositions. \emph{SIAM
Review}, 53(2), 217-288.

\item
Hilafu, H. (2015). Random Sliced Inverse Regression.
\emph{Communications in Statistics-Simulation and Computation},
accepted.

\item
Hilafu, H. and Yin, X. (2017). Sufficient dimension reduction
and variable selection for large-p-small-n data with highly
correlated predictors. \emph{Journal of Computational and
Graphical Statistics}, 26, 26-34.

\item
Lee, M., Shen, H., Huang, J. Z., and Marron, J. S. (2010).
Biclustering via sparse singular value decomposition.
\emph{Biometrics}, 66, 1087-1095.

\item
Li, B., Wen, S. and Zhu, L. (2008). On a projective resampling
method for dimension reduction with multivariate responses.
\emph{Journal of the American Statistical Association},
103(483), 1177-1186.

\item
Li, K. C. (1991). Sliced inverse regression for dimension
reduction. \emph{Journal of the American Statistical
Association}, 86, 316-327.

\item
Li, L., Cook, R. D., and Tsai, C. L. (2007). Partial inverse
regression. \emph{Biometrika}, 94(3), 615-625.

\item
Li, R., Zhong, W., and Zhu, L. (2012). Feature screening via
distance correlation learning. \emph{Journal of the American
Statistical Association}, 107(499), 1129-1139.

\item
Ma, Y. and Zhu, L. (2013). A review on dimension reduction. {\it
International Statistical Review}, 81, 134-150.

\item
Szekely, G. J., Rizzo, M. L., and Bakirov, N. K. (2007).
Measuring and testing dependence by correlation of distances.
\emph{The Annals of Statistics}, 35, 2769-2794.

\item
Tyler, D. E. (1981). Asymptotic inference for eigenvectors.
\emph{The Annals of Statistics}, 9, 725-736.

\item
Woodruff, D. P. (2014). Sketching as a tool for numerical linear
algebra. \emph{Foundations and Trends in Theoretical Computer
Science}, 10(1-2), 1-157.

\item
Yin, X. and Hilafu, H. (2015). Sequential sufficient dimension
reduction for large p, small n problems. \emph{Journal of the
Royal Statistical Society: Series B}, 77, 879-892.

\item
Zhu, L., Miao, B., and Peng, H. (2006). On sliced inverse
regression with high-dimensional covariates. \emph{Journal of
the American Statistical Association}, 101, 630-643.

\item
Zhu, L. P., Li, L., Li, R., and Zhu, L. X. (2011). Model-free
feature screening for ultrahigh-dimensional data. \emph{Journal
of the American Statistical Association}, 106, 1464-1475.

\item
Zhu, L. P., Zhu, L. X., Ferre, L., and Wang, T. (2010).
Sufficient dimension reduction through
discretization-expectation estimation. \emph{Biometrika} 97,
295-304.

\end{description}


\clearpage

\clearpage

\clearpage

\clearpage

\begin{table}
\centering \caption{The leave-one-out classification accuracies of
iRP-SDR ($\widehat B_u$) and PCA-SDR at different envelope sizes $u$
in the EEG data analysis.}\label{tab.eeg} \vspace{3ex}
\begin{tabular}{ccc}
\hline
$u$ &   iRP-SDR &   PCA-SDR \\
\hline
30  &   0.803   &   0.721   \\
35  &   0.844   &   0.787   \\
40  &   0.828   &   0.820   \\
45  &   0.853   &   0.795   \\
50  &   0.844   &   0.779   \\
55  &   0.853   &   0.771   \\
60  &   0.820   &   0.803   \\
\hline

\end{tabular}
\end{table}

\end{document}